\documentclass[5p,times,twocolumn]{elsarticle}
\usepackage{lineno,hyperref}
\usepackage{subcaption}
\usepackage{tikz}
\usepackage{pgfplotstable}
\usepackage{pgfplots}
\usepackage{multirow}
\usepackage{float}
\usepackage{textcomp}
\usepackage{amsmath,amsfonts,amssymb,bbm}	
\usepackage{multicol}
\usepackage{blindtext}

\usepackage[section]{placeins}

\modulolinenumbers[5]

\biboptions{authoryear}

\begin{document}

\begin{frontmatter}

\title{StRegA: Unsupervised Anomaly Detection in Brain MRIs using a Compact Context-encoding
Variational Autoencoder}

\cortext[StRegA]{StRegA: \textbf{S}egmen\textbf{t}ation \textbf{Reg}ularised \textbf{A}nomaly}

\author[1,2,3]{Soumick Chatterjee\corref{mycorrespondingauthor}}
\cortext[mycorrespondingauthor]{Corresponding author:}
\ead{soumick.chatterjee@ovgu.de}

\author[3,4]{Alessandro Sciarra}
\author[1,4]{Max D{\"u}nnwald}
\author[1]{Pavan Tummala}
\author[1]{Shubham Kumar Agrawal}
\author[1]{Aishwarya Jauhari}
\author[1]{Aman Kalra}
\author[4,6]{Steffen Oeltze-Jafra}
\author[3,5,6,7]{Oliver~Speck}
\author[1,2,6]{Andreas N{\"u}rnberger}

\address[1]{Faculty of Computer Science, Otto von Guericke University Magdeburg, Germany}
\address[2]{Data and Knowledge Engineering Group, Otto von Guericke University Magdeburg, Germany}
\address[3]{Biomedical Magnetic Resonance, Otto von Guericke University Magdeburg, Germany}
\address[4]{MedDigit, Department of Neurology, Medical Faculty, University Hospital, Magdeburg, Germany}
\address[5]{German Center for Neurodegenerative Disease, Magdeburg, Germany}
\address[6]{Center for Behavioral Brain Sciences, Magdeburg, Germany}
\address[7]{Leibniz Institute for Neurobiology, Magdeburg, Germany}

\begin{abstract}
Expert interpretation of anatomical images of the human brain is the central part of neuro-radiology. Several machine learning-based techniques have been proposed to assist in the analysis process. However, the ML models typically need to be trained to perform a specific task, e.g., brain tumour segmentation or classification. Not only do the corresponding training data require laborious manual annotations, but a wide variety of abnormalities can be present in a human brain MRI - even more than one simultaneously, which renders a representation of all possible anomalies very challenging. Hence, a possible solution is an unsupervised anomaly detection (UAD) system that can learn a data distribution from an unlabelled dataset of healthy subjects and then be applied to detect out-of-distribution samples. Such a technique can then be used to detect anomalies - lesions or abnormalities, for example, brain tumours, without explicitly training the model for that specific pathology. Several Variational Autoencoder (VAE) based techniques have been proposed in the past for this task. Even though they perform very well on controlled artificially simulated anomalies, many of them perform poorly while detecting anomalies in clinical data. This research proposes a compact version of the “context-encoding” VAE (ceVAE) model, combined with pre and post-processing steps, creating a UAD pipeline (StRegA), which is more robust on clinical data  and shows its applicability in detecting anomalies such as tumours in brain MRIs. The proposed pipeline achieved a Dice score of 0.642±0.101 while detecting tumours in T2w images of the BraTS dataset and 0.859±0.112 while detecting artificially induced anomalies, while the best performing baseline achieved 0.522 ±0.135 and 0.783 ±0.111, respectively. 

\end{abstract}

\begin{keyword}
Anomaly Detection \sep Unsupervised Learning \sep MRI \sep Brain Tumour Segmentation
\end{keyword}

\end{frontmatter}


\section{Introduction}
\label{sec:intro}
Acquiring and analysing magnetic resonance images (MRIs) is an integral part of the work of neuroradiologists. MRI is free from harmful ionising radiation and its ability to show excellent soft-tissue contrast, making it the preferred imaging modality for brain imaging in clinical practice~\citep{hagens2019impact}. MR images are used to examine numerous brain pathologies, and manually analysing or annotating such images is part of the radiologists’ regular duties. Not only such duties are time-consuming and laborious, but they are also error-prone~\citep{brady2017error}. Noting from \citet{bruno2015understanding}, diagnostics of brain pathology remain undiscovered in up to 5-10 \% of cases. The medical image analysis community has made numerous contributions toward automating specific steps, focusing on human-level accuracy at economically adequate computational complexity. With advancements in artificial intelligence, outstanding achievements have been marked in the detection and segmentation of lesions. Furthermore, these scientific results display abilities on par with experts \citep{menze2014multimodal}. Such automated methods try to assist the radiologists in the decision-making process - reducing the workload and observer dependency while improving the workflow and diagnostic accuracy \citep{guerrero2018white}. 

Most of these automatic methods are based on supervised learning techniques, relying on the quality and quantity of labelled data. They utilise a vast amount of annotated training data  -  ranging from thousands to millions of samples. These trained models are usually specialists in dealing with domain-specific tasks or the tasks they have been optimised for - which can also be considered as the bottleneck in such a method as they typically follow a ‘focus point solution’ design, which means they are often very focused on one type of pathology or lesion. Considering an example of cerebral small vessel disease, supervised solutions are challenging to develop since the damage is complex in terms of lesion size, contrast, or morphology~\citep{wardlaw2013neuroimaging}. Considering the complex field of medicine, procuring annotated data is a tedious manual process - while it might be challenging to annotate concerning all the possible pathologies.  

An alternative solution to the conventional approach of the supervised machine learning method to detect brain pathologies is an unsupervised anomaly detection (UAD) system. Anomaly detection is an approach that distinguishes anomalies completely based on characteristics that describe regular data. Since the feature of possible irregularities is not learned, they stand from the normal distribution and can be consequently detected. Anomaly detection solutions are useful in cases where anomalies are to be detected, but their manifestation is not known, and they even might occur very rarely - making it difficult to create a large dataset for supervised training~\citep{pimentel2014review}. Anomaly detection techniques have been applied for multiple tasks, including credit card fraud detection \citep{phua2004minority}, aerospace monitoring during flight \citep{clifton2007framework}, illegal object detection at airports \citep{akcay2018ganomaly} or detecting faulty microprocessors \citep{kim2012machine}. Approach without using deep learning, by exploiting representative neighbours - ADRN, has also been proposed for the detection of anomalies in different datasets~\cite{liu2021anomaly}.

Anomaly detection in medical imaging is defined as identifying relevant indicators of diseases and differentiating them from those of typical healthy tissue characteristics. UAD systems can directly learn the data distribution from a large cohort of un-annotated healthy subjects to detect out-of-distribution or anomalous samples to identify cases for further inspection. As anomaly detection does not depend on reference annotations, this approach does not require any human input and can therefore be used to identify any medical condition. Unsupervised segmentation of brain regions for anomaly detection is clinically relevant as it can assist radiologists. It is also of great interest to the medical image analysis community as it eliminates the need for pixel-level annotations, which is an expensive process and is often a prerequisite for supervised training of the networks. However, it is worth mentioning that a system, such as UAD, does not differentiate between different types of anomalies, i.e. lesions and artefacts - both will be treated as anomalies by a UAD system. Variational autoencoders (VAEs) are one of the most commonly used techniques for anomaly detection, along with methods based on Generative Adversarial Networks (GANs)~\citep{zimmerer2019unsupervised,liu2020using,garcia2021usage,pinaya2022unsupervised}. However, many of such techniques fail to perform well when it comes to real clinical data. This research aims to extend the research field of anomaly detection by creating a UAD pipeline capable of working with clinical data for the detection of clinically-relevant anomalies (e.g. lesions) and evaluating its performance for the task of brain tumour detection. 

\subsection{Contributions}
This paper presents an unsupervised anomaly detection pipeline, StRegA - Segmentation Regularised Anomaly, which combines the proposed compact ceVAE (cceVAE) with pre-processing steps to simplify the input to the model and post-processing steps to improve the prediction of the model. The proposed method has been trained on anomaly-free brain MRI datasets and then was evaluated on the task of brain tumour detection on the T1ce and T2 MRIs from the BraTS dataset. The method was compared against three baseline models, including the original ceVAE, and an additional baseline by combining the original ceVAE with the pre-processing techniques of StRegA. The methods were also evaluated on a synthetically created anomalous dataset. Additionally, this paper also provides a comprehensive overview of different autoencoder-based techniques for anomaly detection.

\section{Related Work}
\label{sec:relatedwork}
Auto-encoders are trained by minimising its reconstruction error~\citep{alain2014regularized}. Different types of reconstruction errors are commonly used in unsupervised anomaly detection, acknowledging that regular distribution models cannot effectively reproduce anomalies~\citep{ribeiro2018study}. Therefore, regular anomaly-free data is distinguishable from anomalous data with the degree of error differences in terms of reconstruction. The majority of the methods assume reconstruction errors (losses) are optimised for average distribution data, whereas anomalies are impacted with significant losses when their dimensions are reduced. 

Generative Adversarial Networks (GANs)~\citep{goodfellow2014generative} and Variational Autoencoder (VAEs) \citep{kingma2013auto}  are two essential methods based on latent space information modelling, and they have been employed for the task of anomaly detection~\citep{xu2018unsupervised,yan2020unsupervised}. However, both these methods have different approaches for modelling the distribution. VAE  applies variational inference to approximate the data distribution. The posterior distribution is encoded into latent space using the encoder network of the VAE, whereas the decoder network is responsible for modelling the likelihood. Unlike VAE, the generator network of the GAN transforms the prior distribution of the random samples in the latent space into data samples that an optimised classifier cannot distinguish. The obtained data distribution is inferable, and GAN is mainly a sampler. Recently some researchers \citep{carrara2021combining} have coupled an autoencoder with GAN to derive enhanced latent representations in the domain of anomaly detection. 

On the other hand, after revolutionising the field of natural language processing, Transformers~\citep{vaswani2017attention} have found applications in other areas such as computer vision, including modelling brain images with impressive results. The defining characteristic of a transformer is that it relies on attention mechanisms to capture
interactions between inputs, regardless of their relative position to one another. The output is computed as a weighted sum of the values, where the weight assigned to each value is computed by a compatibility function of the query with the corresponding key. Transformers have been used to model brain images for the purpose of anomaly detection after assuming an arbitrary ordering to transform the latent discrete variables $z_q$ into a 1D sequence s and training the transformer to maximise the training data’s log-likelihood in an autoregressive fashion~\citep{pinaya2022unsupervised}. However, such methods are typically memory-hungry - making them difficult to use with limited computing resources. VAEs can be less computationally expensive than transformers, making them more suitable for using them widely. Hence, this paper focuses is on VAE-based techniques and some of the related AE and VAE based works which were studied and  experimented with for baseline selection, are discussed in detail.

\subsection{Adversarial Autoencoder}
\label{sec:AdvAE}
Autoencoders~\citep{rumelhart1985learning,kramer1991nonlinear} try to learn an underlying function - which is an approximation to the identity function - trying to reproduce the input as its output, learning a latent space representation (LSR) of the input in the process~\citep{ng2011sparse}. The model's output is compared against the input with the help of reconstruction loss, which is then backpropagated to train the network. Adversarial autoencoder \citep{makhzani2016adversarial} converts an autoencoder into a generative model, which has also been used for the task of anomaly detection \citep{beggel2019robust,li2019video}. The objective is two-fold - a regular reconstruction loss and an adversarial training loss. The second loss criterion tends to match the aggregated posterior distribution of the latent space of the autoencoder to an arbitrary prior distribution. An outcome of this architecture bounds the encoder module to learn the data distribution close to the prior distribution, and the decoder maps the imposed prior to the data distribution. The encoding function of the encoder is formally represented as 
\begin{equation} 
q(z) = \int_x q(z\mid x) p_d(x) dx 
\end{equation} 
where $x$ is the input and $z$ is its latent representation vector. $p(z)$ be the prior distribution to be imposed on the latent vectors, $q(z|x)$ is an encoding dimension, and p(x$\mid$ z) is the decoding distribution. Also, $p_d(x)$ is the data distribution and $p(x)$ is the model distribution. The adversarial autoencoder is regularised by matching this aggregated posterior $q(z)$ to an arbitrary prior, $p(z)$. The whole training takes place in two phases - \emph{reconstruction} phase where the encoder and decoder are updated to minimise the reconstruction error in the input. Later in the \emph{regularisation} phase, the adversarial network updates the discriminator to classify better actual samples (from the prior) from the generated samples (from the encoder). Finally, the generator or the encoder is updated to confuse the discriminator. 

\subsection{Scale-Space Autoencoder}
\label{sec:ssAE}
Many variations of autoencoders have been proposed. One of them is  Scale-Space Autoencoder (SSAE) \citep{baur2020scalespace}.This framework is used for unsupervised anomaly detection based on the Laplacian pyramid, allowing efficient compression and reconstruction of brain MRI images with high fidelity and successfully suppressing anomalies. Contrary to other frameworks, the SSAE framework models the distribution of the scale-space representation of healthy brain MRI. The Laplacian pyramid is a multi-scale representation of the input image; it allows efficient segmentation of anomalies at different resolutions and then aggregates the results. SSAE architecture is an encoder-decoder network that aims to localise anomalies from reconstruction residuals. The frequency band of the input data is split by compressing and reconstructing the Laplacian pyramid of the healthy brain MRI. The input image $x$ is repeatedly smoothed and downsampled to obtain a Laplacian pyramid with K levels. Reconstruction can be recursively determined using
\begin{equation}
   \hat{x} = \sum_{k=0}^{K-1} u(I_{K-k}) + H_{K-1-k}
\end{equation}
where $I_k$ is the low-resolution representation of the image x after K down samplings and high-frequency residuals $H_0,..,H_{K-1}$ which are obtained at each level k using :
\begin{equation}
   H_k = I_k - u(I_{k+1}) ; \forall 0	\leq k<K
\end{equation}
The overall loss can be defined as the weighted sum of losses at all scales:
\begin{equation}
   L = \sum_{k=0}^{K} \lambda_kL_k
\end{equation}
where,
\begin{equation}
   L_k = l2_{norm}(I_k, \hat{I_k}) = l2(I_k, u(\hat{I_k}+\hat{H_k})
\end{equation}

\subsection{Bayesian Skip autoencoder}
\label{sec:SkipAE}
Bayesian Skip autoencoder  \citep{cccc}  is formulated on a concept that has been used and proven advantageous for biomedical image segmentation. This framework makes use of skip connections that allow the models with limited capacity to be trained and utilised for unsupervised anomaly detection. Skip connections in autoencoders facilitate the reconstruction of high-resolution and complex medical data with high fidelity. A drop-out-based identity mitigation technique is employed to prevent the model from learning an identity mapping and allowing it to bypass the lower levels of the autoencoder. The loss function for the model can be defined as:
 \begin{equation}
   L(x, \hat{x}) = L1(x, \hat{x}) - gdl(x, \hat{x})
\end{equation}
Here, $L1$  is defined as the L1 Manhattan distance between the input and its reconstruction $x$ and $\hat{x}$ respectively, $gdl(.,.)$ is the gradient difference loss.

\subsection{Variational Autoencoder (VAE)}
\label{sec:VAE}
Variational autoencoders (VAEs)~\citep{kingma2013auto} belong to a class of generative models that provides a probabilistic way to represent the observations in the latent space. VAE and its flavours are one of the first and most popular models for unsupervised anomaly detection~\citep{an2015variational,xu2018unsupervised} Unlike an autoencoder, the encoder of the VAE describes the probability distribution of each latent attribute. During training, the model computes a latent space representation (LSR) of the non-anomalous input image. The latent space in VAEs is continuous by design and allows easy random sampling and interpolation. This is achieved by designing the encoder to output two vectors:  a vector of means $\mu$ of data samples and a vector of standard deviations $\sigma$. While working with images, the $\mu$ and $\sigma$ are calculated from the intensity values of the encoded images. The encoding is generated within these distributions. The decoder learns that not just a single point but all the points around it are referred to as a sample of the same class, thus decoding even slight variations of the encoding. Typically, the network is trained by optimising the evidence lower bound (ELBO) $L$, which can be defined as:

\begin{equation}
\log p(x) \geq L=-D_{K L}(q(z \mid x) \| p(z))+E_{q(z \mid x)}[\log p(x \mid z)]
\end{equation}
where, q(z$\mid$ x) and  p(x$\mid$ z) are diagonal normal distributions that are parameterised by neural networks $f_\mu$, $f_\sigma$, and $g_\mu$, and constant $c$ such that:

\begin{equation}
\mathrm{q}(\mathrm{z} \mid x)=N\left(z ; f_{\mu, \theta 1}(x), f_{\sigma, \theta 2}(x)^{2}\right)
\end{equation}
\begin{equation}
p(x \mid z)=N\left(x ; g_{\mu, \gamma}(z), I * c\right)
\end{equation}
 
\subsection{Vector  Quantized  Variational  autoencoder}
\label{sec:VQVAE}
Vector Quantised-Variational AutoEncoder (VQ-VAE) \citep{oord2017neural,marimont2020anomaly} framework encodes the images in categorical latent space. Further, an auto-regressive (AR) model is used to model the prior distribution of latent codes. The prior probability approximated by the AR model is then used for unsupervised anomaly detection and gives the sample and pixel-wise anomaly scores. The encoder of the VQ-VAE framework is based on a dictionary mapping K discrete keys into a D-dimensional embedding space, i.e., observed variables are mapped to the embedding space, $z_e(x) \in R^D$. The posterior in VQ-VAE is defined as :
\begin{equation}
    q(z = k \mid x) = 
    \begin{cases}
            1, & for k = \operatorname*{arg\,min}_j || z_e(x) - e_j ||_2\\
    0, & \text{otherwise}
    \end{cases}
\end{equation}
The embedding in the dictionary nearest to the encoder's output is given to the decoder as an input, i.e.,$z_q(x) \sim q(z\mid x)$. The decoder then reconstructs the distribution $p(x\mid z_q(x))$.The VQ-VAE Loss function is therefore defined as:
\begin{equation}
  L = log(x\mid z_q(x)) + \mid\mid sg[z_e(x)]-e\mid\mid_2^2 + \mid\mid sg[e] - z_e(x)\mid\mid_2^2  
\end{equation}
where $sg[.]$ is the stop gradient operator.
Further, an AR model is employed to learn the prior distribution of VQ-VAE, and the probability of samples is approximated. The samples with low probability are identified as anomalies. AR models generate multiple restorations as they are generative models and allow sampling of one variable at a time in an iterative manner. 

\subsection{Gaussian  Mixture  Variational  Autoencoder}
\label{sec:GMVAE}
The Gaussian  mixture  variational  autoencoder \citep{dilokthanakul2016deep,chen2020unsupervised}  approach (GMVAE) combines image restoration techniques with unsupervised learning - used in various anomaly detection approaches~\citep{guo2018multidimensional,fan2020video}. A GMVAE is first trained on an anomaly-free dataset to learn a normative prior distribution. A Maximum-A-Posteriori (MAP) restoration model then uses this prior to detecting the outliers. In GMVAE, the ELBO is formally expressed as:
    \begin{equation}
    \begin{aligned}
     \mathbb{E}{q(z\mid X)} [log p(X\mid z)] - \mathbb{E}{q(\omega \mid X) p(c \mid z, \omega)} \\
             [KL [q(z \mid X) | p(z \mid \omega, c)]]\\
             - [KL [q(\omega \mid X) | p(\omega)]] - \mathbb{E}_{q(z \mid X) q(\omega \mid X)} \\
             [KL [p(c \mid z, \omega) | p(c)]]
    \end{aligned}
    \end{equation}
            
where the first part is the reconstruction term, the second part ensures the encoder distribution fits the prior, the third term does not let the posterior $\omega$ diverge from the prior $p(\omega)$, and the last term enforces the model to not collapse into a single Gaussian but uses the mixture model. For the architectural choice, seven convolutional layers for the encoder and seven transpose layers for the decoder are used. The latent variables $z$ and $\omega$ are 2D structures with size (32, 42 1) and nine clusters, c.

\subsection{Context-encoding Variational Autoencoder}
\label{sec:ceVAE}
The core idea of Context-encoding Variational Autoencoder (ceVAE) \citep{zimmerer2018context} is to combine the reconstruction term with a density-based anomaly scoring. This allows using the model-internal latent representation deviations and an expressive reconstruction term on the samples and individual pixels. The overall architecture comprises of fully convolutional encoders $f_u$, $f_\sigma$ and a decoder unit $g$, where the context encoding part used the encoder module for a data sample. ceVAE can be divided into two branches: VAE branch and CE branch.

\paragraph{\textbf{VAE Branch}:} The deviations of the latent from the CE branch is further inspected. The encoders $f_u, f_\sigma$, the decoder g, and a standard diagonal Gaussian prior p(z) are used in the VAE module with an objective function:
    
    \begin{equation}
    L_{VAE} = L_{KL} (f_\mu(x), f_\sigma(x)^2) + L_{recVAE}(x, g(z))
    \end{equation} 
    
    where $z \sim \mathcal{N}(f_\mu(x), f_\sigma(x)^2)$ with the reparameterisation trick and $L_{KL}$ loss with a standard Gaussian yield a comparable anomaly score. KL-loss is used to analyse the deviations of posterior from the prior. 
    
    \paragraph{\textbf{CE branch}:} sample $x$ is subjected to context encoding noise by masking specific regions. Training occurs by reconstructing the perturbed input $\hat{x}$ using the encoder $f_\mu$ and the decoder g, formally, $L_{rec CE}(x, g(f_\mu(\hat{x})))$. The results show CEs are more discriminative and semantically produce rich representations, which positively influences the expressiveness of model variations. 
    
    \paragraph{\textbf{ceVAE}:} On combining the above modules, the aim is to capture both the effects, i.e., a better-calibrated reconstruction loss and model-internal variations. The objective function is restructured as the following: 
    
    \begin{equation}
    \begin{aligned}
    L_{ceVAE} = L_{KL}(f_\mu(x), f_\sigma(x)^2) \\ 
                + L_{recVAE}(x, g(z)) + L_{recCE}(x, g(f_\mu(\hat{x})))
    \end{aligned}
    \end{equation}
    
    where $L_{KL}$ is the KL-loss, z is sampled with the reparameterisation method, and $\hat{x}$ is perturbed by masking regions similar to CEs. During the training process $L_{recCE}$ has no constraints for normality on the prior $p(z\mid x)$. The combination of CE and VAE has a regularising effect that prevents the posterior from collapsing and better represents the semantics in the input data.


\subsection{Hypothesis behind StRegA}
Even though VAEs have their advantages, like they can can be less computationally expensive than transformers, but have been seen to struggle to detect anomalies in clinical data. The reason for this could be the differences between the training and testing sets in terms of image properties (e.g. contrast). One possible way to improve their performance would be to "simplify" the input to the model - making it easier for the model to learn a distribution of the anomaly-free data - which might also help the model generalise better. Hence, this paper focuses on developing a VAE-based technique focusing on output "simplification" to improve the model's performance on clinical data.

\section{Methodology}
\label{sec:methods}

\subsection{Study of existing methods and baseline selection}
Autoencoders and their flavours are commonly used for unsupervised anomaly detection as discussed in~\ref{sec:relatedwork}. While selecting the baselines, the aforementioned methods were compared and finally, the best performing baseline (i.e. ceVAE) was selected and modified to develop the proposed approach.
\subsection{StRegA : Segmentation Regularised Anomaly }
\begin{figure*}
    \centering
    \includegraphics[width=0.95\textwidth] {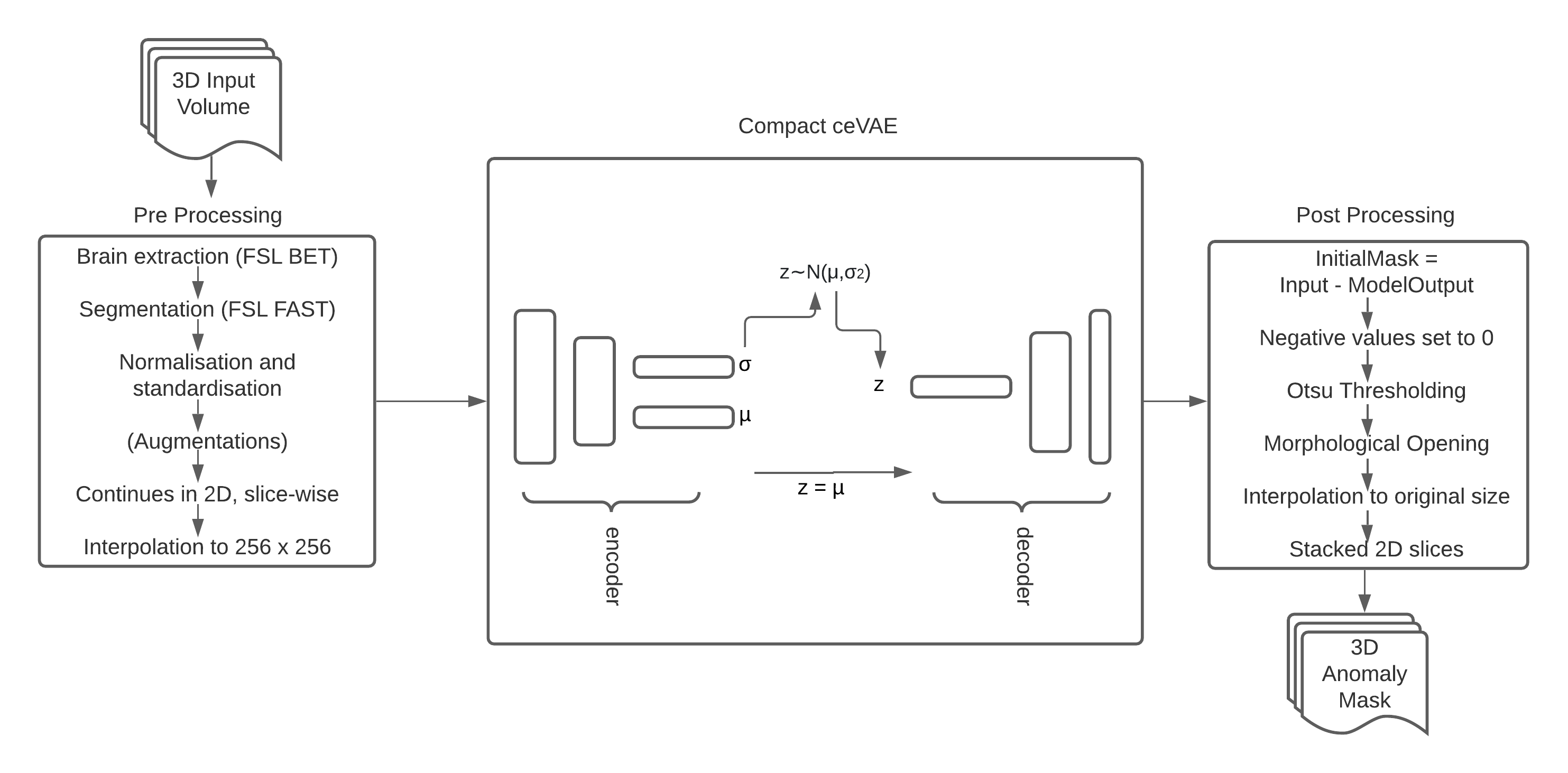}
    \caption{StRegA pipeline} 
    \label{fig:StRegA}
\end{figure*}

The proposed unsupervised anomaly detection pipeline, StRegA: Segmentation Regularised Anomaly - shown in Fig.\ref{fig:StRegA}, includes a modified compact version of the Context-encoding Variational Autoencoder (ceVAE) \citep{zimmerer2018context} - Compact ceVAE (cceVAE), combined with pre and postprocessing steps. Anomaly detection is performed in 2D, per-slice basis. The input volumes are preprocessed in 3D, then slices are supplied to the model, followed by that, postprocessing steps are applied, and finally, the slices are stacked together to obtain the final result in 3D. It is worth mentioning by splitting the 3D volumes into 2D slices, the spatial context within the volumes is lost. But this was done to make the training process feasible in terms of GPU requirements. The code of StRegA is publicly available on GitHub~\footnote{StRegA on GitHub: \url{https://github.com/soumickmj/StRegA}}.

\subsubsection{Preprocessing}
\label{sec:Preprocessing}
The first preprocessing step was to segment the input image into different tissue types and obtain the corresponding segmentation mask. Segmentation means partitioning the input images into various segments. Segmentation can simplify the input given to the model as the model needs to deal with less variations in the input images that can help improve the performance of the anomaly detection system. Segmentation helps standardise the input images with different image properties, such as images with different MR contrasts (e.g. T1w, T2w). This was performed by supplying the 3D brain images to the automatic segmentation tool of FSL (\cite{jenkinson2012fsl}) - FAST, which segmented the brain into three different tissue types: grey matter, white matter, cerebrospinal fluid (CSF); and background - a total of four classes. The segmentation command of FSL FAST has a flag to select the modality of the input volume (e.g. T1w, T2w). Rest of the parameters of FSL FAST (apart from the number of classes and the type of the image) were kept to their default values. The underlying principle of FAST is formulated on a hidden Markov random field model and an associated expectation-maximisation algorithm. The entire process is automated and is capable of producing a probabilistic/partial volume tissue segmentation. Apart from segmenting the input volume, this tool also corrects spatial intensity variations (also known as bias field or RF inhomogeneities). The segmentation mask was normalised with the intensity distribution of the dataset - normalised with zero mean and unit variance. The volumes were divided into 2D slices in the axial orientation, and then they were resized to 256×256 in-plane using bilinear interpolation before feeding them into the network. During the training phase, in addition to the above-mentioned preprocessing steps, data augmentation was also performed. 

\paragraph{Data Augmentation} The augmentation techniques used can be split into two groups: intensity-based and spatial augmentation. Intensity augmentation included bias field artefacts - where the bias field is modelled as a linear combination of polynomial basis functions, as in \cite{van1999automated} with the maximum magnitude of polynomial coefficients set to 0.5 and order of the basis polynomial functions to 3, Gaussian noise with the standard deviation of the noise distribution chosen randomly in a uniform fashion between 0 and 0.25, random gamma with gamma value being uniformly-chosen between -0.3 and +0.3, and ghosting artefacts with the number of ghosts being uniform between 4 and 10. On the other hand, spatial augmentation consisted of horizontal flipping, vertical flipping, affine transformations (range of degrees from -35 and +35, using nearest neighbour interpolation), and rotation with a degree between -15 and +15.



\subsubsection{Network Model: cceVAE}
\label{sec:model}

\begin{figure*}
    \centering
    \includegraphics[width=0.95\textwidth] {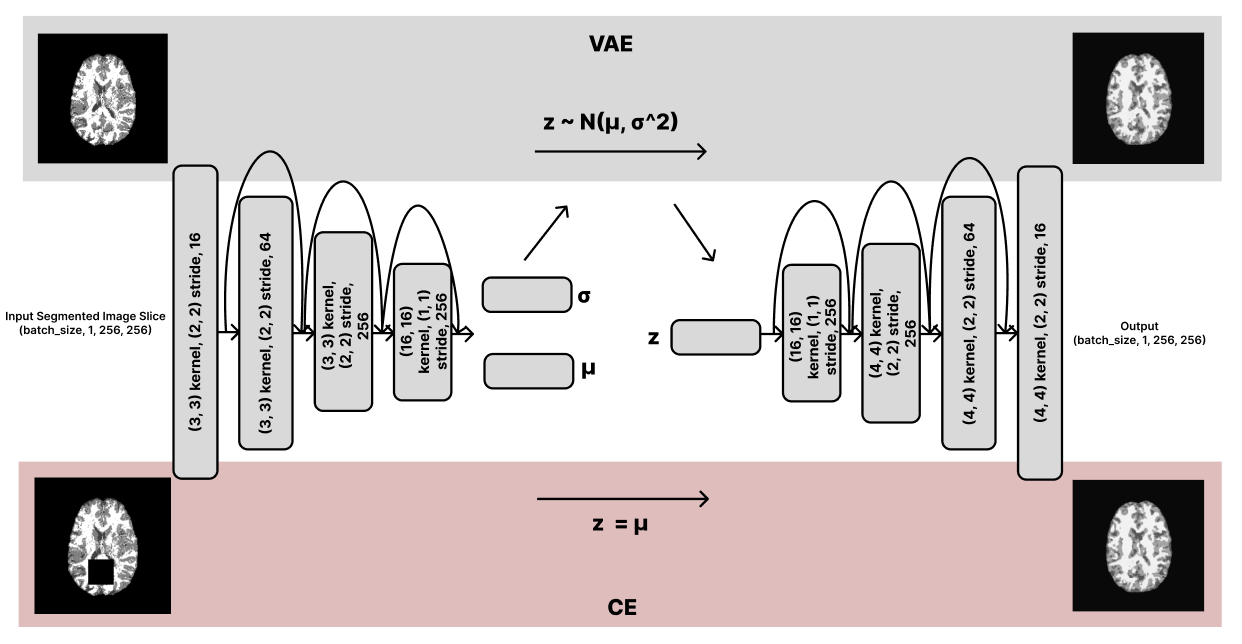}
    \caption{The proposed Compact ceVAE architecture: cceVAE. Each of the convolution blocks contains a convolution layer with the mentioned configuration, followed by a batch normalisation layer.} 
    \label{fig:cceVAE}
\end{figure*}

This work proposes a compact modified version of the ceVAE model \citep{zimmerer2018context}, explained in Sec.\ref{sec:ceVAE}: Compact ceVAE (ccVAE). A representational diagram of the network has been shown in Fig.~\ref{fig:cceVAE}. This is a compact version as it has smaller encoder and decoder than the original ceVAE - having a symmetric encoder-decoder with 64, 128 and 256 feature maps and a latent variable size of 256. The authors hypothesise that a more compact version of the network will reduce any possibilities of overfitting, especially for data that is more simplified than un-processed slices - where many learnable parameters might not be required. This proposed model also uses residual connections and batch normalisation layers - which were not present in ceVAE. The preprocessed images are fed into this network model, and the network tries to reconstruct the input. During training, anomaly-free (in the experiments of this paper: healthy brain MRIs) images are provided, and the network attempts to learn a Gaussian distribution of the given anomaly-free dataset. While inference, the network attempts to reconstruct the input image from the learnt Gaussian distribution. If the network encounters images with anomalies - they are then out-of-distribution samples. Hence, the network should "in-theory" fail to reconstruct the anomalies as they do not belong to the learnt distribution. The reconstruction difference is then computed by subtracting the model's output from its input to generate the initial anomaly mask - which is then postprocessed to finally detect the anomaly.

\subsubsection{Postprocessing}
\label{sec:Postprocessing}
After subtracting the reconstruction from the model's input, the initial anomaly mask is obtained - containing values between zero and one, where one implies anomaly while zero means no anomaly. Then this mask is first thresholded to eliminate any pixel having a value less than zero by replacing all the negative values with zero. Then to remove the intensity outliers, the mask was thresholded using Otsu's method \footnote{\url{https://en.wikipedia.org/wiki/Otsu's_method}}. The Otsu thresholding was applied individually on each output slice - resulting in different suitable thresholds for each slice of each subject. Otsu is a frequently used method for performing automatic image thresholding, which divides all the pixels into two classes - background and foreground, based on a single intensity threshold. The threshold is calculated by minimising intra-class intensity variance, or in other words, by maximising inter-class variance. \footnote{\url{http://www.labbookpages.co.uk/software/imgProc/otsuThreshold.html}}. 

This results in a binary anomaly mask. After performing Otsu thresholding, the small irrelevant parts of the mask are removed by applying morphological opening \footnote{\url{https://en.wikipedia.org/wiki/Opening_(morphology)}} - which removes all bright structures of an image with a surface smaller than the area threshold. It is equivalent to applying dilation on top of erosion on the image. This gives the final anomaly mask. It is then resized through interpolation to whichever input size the original image had before resizing during preprocessing. Finally, all the slice-wise 2D masks are stacked together to get the final anomaly mask in 3D.


\subsection{Datasets}
\label{sec:Datasets}
Training unsupervised anomaly detection models warrant anomaly-free datasets. As the focus of this paper is on brain MRIs, two anomaly-free datasets were selected for training the models: MOOD challenge dataset \citep{mood_2021_4573948} and IXI dataset \footnote{IXI Dataset: \url{ https://brain-development.org/ixi-dataset/}}. It is worth mentioning that even though both the datasets contain MRIs only from normal, healthy subjects - there might still be some anomalies present in some of the MRIs that the experts missed. However, in this research, following the authors of the datasets - these datasets were considered to be comprised of only anomaly-free healthy brain MRIs. MOOD dataset comprises 800 T1w brain MRIs with a matrix size of 256x256x256 - 700 of which were used for training, and 100 were held out for testing, while the IXI dataset contains MRIs of nearly 600 subjects - all of which were used in training, acquired using 1.5T and 3T MRIs in three different hospitals, with different MRI sequences: T1w, T2w, PDw, MRA, and DTI. In this research, T1w and T2w images from the IXI dataset were used separately - for training separate T1 and T2 models. Trainings were performed by combining MOOD with one of the IXI contrast: MOOD+IXIT1 and MOOD+IXIT2. Due to the fact that the models worked with segmented volumes (see Sec.~\ref{sec:Preprocessing}, it was possible to combine MRIs with different contrasts in the training set as the segmentation helps to standardise them. However, the segmentation still has some minute differences in the different contrasts, especially in terms of the tumour delineation. Hence, the motivation behind combining different datasets (and contrasts) was to make the model more robust against inter-dataset variations (e.g. resolution), including differences in contrasts in the test set than in the training set.

For evaluating the approach with a dataset with clinical anomalies, 80 randomly selected volumes from BraTS 2017 \footnote{BraTs: \url{https://www.med.upenn.edu/sbia/brats2017/data.html}} dataset was used, which contains four types of MRIs with high and low-grade brain tumours: T1-weighted, and contrast-enhanced T1-weighted (with contrast agent), T2-FLAIR, and T2-weighted. This paper used contrast-enhanced T1-weighted (T1ce) and T2-weighted MRIs. Additionally, the model was also tested on the "toy test set" provided in the MOOD dataset, comprising one anomaly-free and three anomalous MRIs containing artificial anomalies. It is noteworthy that the image contrasts, acquisition parameters, and other image properties of the BraTS dataset are very different from the training sets. Hence, a controlled synthetic anomalous dataset was created as a part of this research - to evaluate the approach on a dataset that contains anomalies, but the contrast and other image characteristics are similar to the training set. 

\subsubsection{Generating Synthetic Anomalous Dataset}
\label{sec:Synthetic_Datasets}
\begin{figure}
    \centering
    \includegraphics[width=0.48\textwidth] {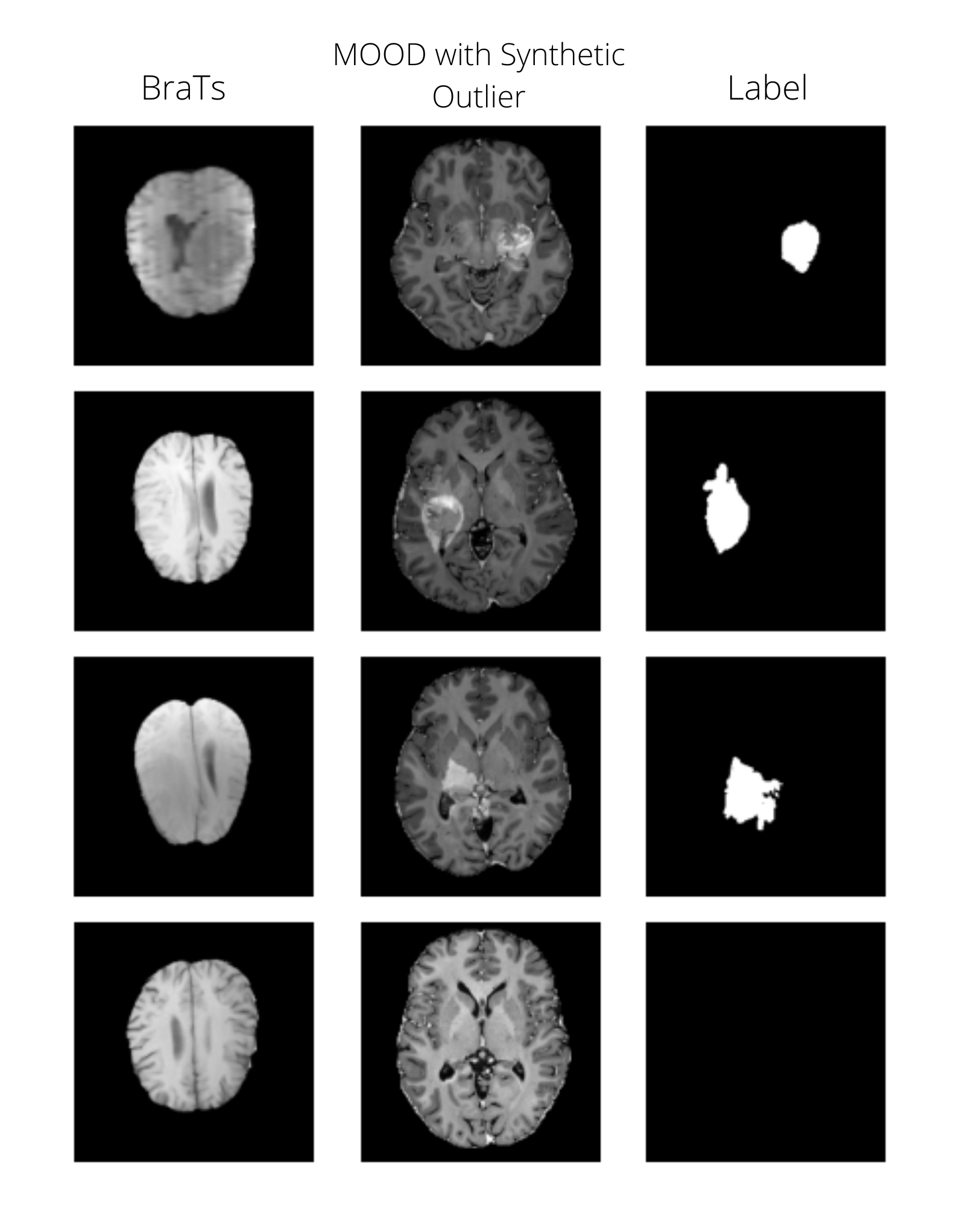}
    \caption{Synthetic MOOD image generated using anomaly from BRATS data set, including one anomaly-free example in the last row}
    \label{fig: syn}
\end{figure}
An artificial controlled test set was created by superimposing anomalies from the BraTS dataset onto the held-out test-set containing anomaly-free brain MRIs. This was performed to evaluate the performance of the methods when the non-anomalous parts of the images have similar contrast and other characteristics as the training dataset - to solely evaluate the performance of the model when the model doesn't encounter any other difficulties, while it also ensures that the artificial "tumours" are realistic.  

To generate the synthetic anomalous dataset, the anomalies (i.e. tumours) from 20 volumes of the BRATS dataset were extracted by multiplying the ground truth tumour masks with the brain images. Then the intensity values of the tumours were normalised between 0 and 1, and the resultant outputs were interpolated as 256x256 slices. These 20 extracted tumours were superimposed onto 20 randomly-selected volumes from the anomaly-free held-out test-set from the MOOD dataset, which places the tumours on top of the anomaly-free brain images. This creates a set of synthetic anomalous test data which can be used in conjunction to test the performance of the pipeline. Consider a BraTS image sample B, and its corresponding ground-truth segmentation mask S from the dataset pointing out the tumour, then the superimposed MOOD image (from the held-out test-set) sample A of corresponding slice from h slices can be calculated as :

\begin{equation}
  \begin{array}{l}
    A'_{i} = A_{i} + M_{i}\ ; \forall i \in h \text{ ,where }    \\ \cr
     M_{j} = B_{j}\ .\ S_{j}\ ; \forall j \in h
  \end{array}
\end{equation}


\section{Experiments and Evaluations}
\label{sec:results}
The methods were compared based on their accuracy in segmenting the anomalies, calculated with Sørensen–Dice coefficient. Initial experiments were performed using a vanilla VAE~\citep{mickISMRM21Anomaly}, following that, different state-of-the-art models (the ones mentioned in Sec.~\ref{sec:relatedwork}) were compared. Finally, the three best-performing models in this current experimental setup were chosen as baselines, and final comparisons were performed against the proposed method.

\subsection{Initial Selection of Baseline Models}
The initial approach of using the vanilla VAE~\citep{mickISMRM21Anomaly} was able to detect simple contrast anomalies - anomalies having very different contrast than the rest (e.g. MOOD toy dataset), but in most cases failed to generate an acceptable reconstruction. SkipAE~\citep{cccc} resulted in better detection of anomalies. While the results on the synthetic data and the MOOD toy test data were better, the more complicated anomalies were harder to detect in the reconstructions. Other models, which are explained in Sec.~\ref{sec:relatedwork} were also experimented with, and it was found that the ceVAE~\citep{zimmerer2018context}~(Sec.~\ref{sec:ceVAE}) performed the best when it comes to segmenting pathological anomalies. Hence, this was chosen as the primary baseline of this paper. The second and third best performing models, GMVAE~\citep{chen2020unsupervised}~(Sec.~\ref{sec:GMVAE}) and SkipAE~\citep{cccc}~(Sec.~\ref{sec:SkipAE}) respectively, were also chosen as additional baselines. 

\subsection{Method Development}
\label{sec:method_Dev}
\begin{figure}
    \centering
    \begin{subfigure}[b]{0.45\textwidth}
         \centering
         \includegraphics[width=\textwidth] {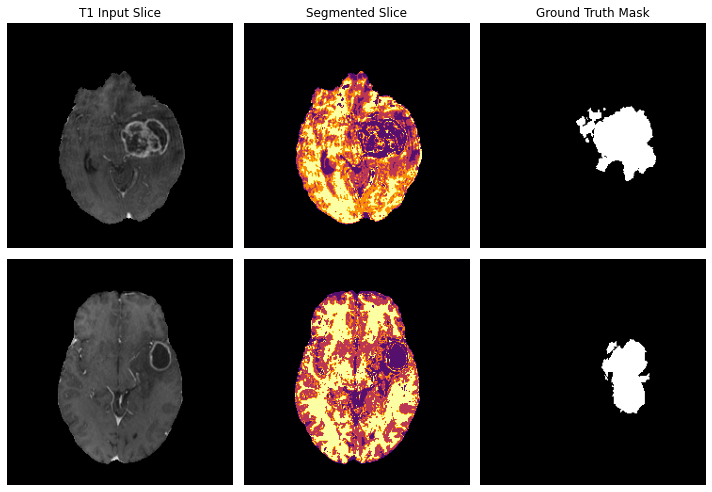}
         \caption{T1w}
         \label{fig:fastT1}
     \end{subfigure}
     \hfill
     \begin{subfigure}[b]{0.45\textwidth}
         \centering
         \includegraphics[width=\textwidth] {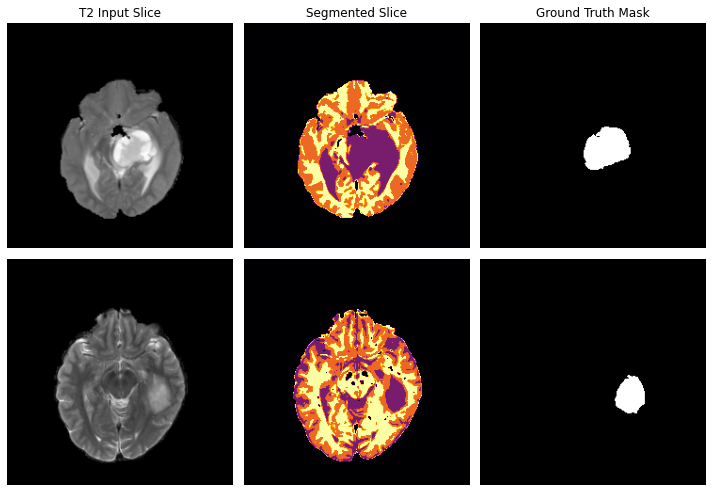}
         \caption{T2w}
         \label{fig:fastT2}
     \end{subfigure}
    
    \caption{Examples of segmented T1w and T2w volumes using FSL FAST}
    \label{fig:fastseg}
\end{figure}

This research focused on simplifying the input to the backbone model - by introducing different preprocessing techniques. It was observed that using segmentation as a preprocessing step assisted the models most and improved the model's performance which was tested with all the baseline models. Fig.~\ref{fig:fastseg} shows a few examples of T1w and T2w slices with tumours applying FSL FAST - to visualise how this preprocessing segmentation performs while working with the BraTS dataset that contains tumours. It was observed that the tumour was usually made part of the CSF class, however, lacks proper delineation. Apart from the tumour, the actual CSF is also part of this class - making it impossible to use these as final tumour segmentations. When this particular class was compared against the corresponding tumour masks, the resultant Dice scores were 0.196±0.116 and 0.202±0.137 for T1w and T2w volumes from the BraTS dataset. Hence, it is required to further proceed with the anomaly detection pipeline to finally localise and segment the tumour. Moreover, as stated earlier, the segmentation step helps to make the MRIs of different contrasts similar - that can also be observed in Fig.~\ref{fig:fastT1}~and~\ref{fig:fastT2}.

Different other preprocessing techniques were also experimented with, and the final set of techniques was made part of the StRegA pipeline - explained in Sec.~\ref{sec:Preprocessing}. When experimenting with the number of layers in the encoder and decoder of the ceVAE model with these preprocessing techniques, it was observed that the results on the test datasets improved significantly when a more compact model was used. This could be due to the fact that a more compact model reduces the risk of over-fitting on the relatively simpler segmented data. Thus after several experiments, having symmetric 64, 128 and 256 feature maps and a latent variable size of 256 yielded the best generalisation in reconstruction quality. Implementing skip connections as was used in SkipAE provided no meaningful improvements, but residual connections with batch normalisation provided better results and were thus included in the model - results in the final cceVAE model~(Sec.~\ref{sec:model}).%
For the experiments here, the proposed model was trained with the Adam optimiser~\citep{kingma2014adam} using a learning rate of 1e-4 and a batch size of 64 for 70 epochs.

\subsubsection{Comparative Results}

\begin{figure*}
        \centering
        \includegraphics [width=0.9\textwidth] {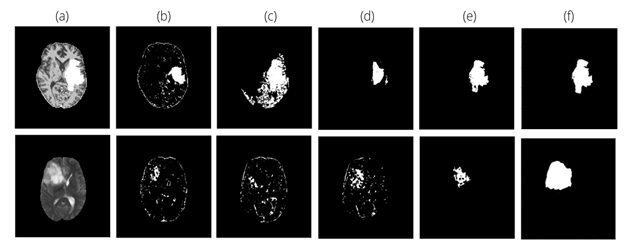}
        \caption{Comparative results from different methods. Top row: artificially generated anomalous data. Bottom row: BraTS T2w image. (a) A slice from the input volume, (b) ceVAE, (c) SkipVAE, (d) GMVAE, (e) StRegA and (f) ground truth} 
        \label{fig:compare_res}
\end{figure*}

The proposed StRegA pipeline was compared against the three baseline models and an additional baseline - by combining the StRegA preprocessing steps with the ceVAE model. Fig.~\ref{fig:compare_res} shows a qualitative comparison of the different methods. It can be observed that the proposed StRegA pipeline provides a more localised detection of the anomaly compared to the other methods. Tables~\ref{tab:t1comparenumbers}~and~\ref{tab:t2comparenumbers} portray a quantitative comparison of the T1 and T2 models. The proposed method outperformed all the other models with statistical significance in both T1 and T2 cases while achieving 49\% and 82\% improvements in Dice scores over the baseline ceVAE for the task of segmenting tumours from T1w and T2w brain MRIs from the BraTS dataset, resulting in average Dice scores of 0.531±0.112 and 0.642±0.101, respectively. A separate model was also trained only on the IXI T2w dataset and was then tested using the BraTS T2w dataset, resulting in an average Dice score of 0.631±0.117 which is 1.7\% less than the model trained on a combination of MOOD (T1) and IXI T2. This can bee attributed to the fact that this specific training had a smaller training set than the combined one. Another additional evaluation was performed by applying the MOOD-IXIT2 traine StRegA on the BraTS T1 images, and it resulted in an average Dice score of 0.533±0.179 (similar to BraTS T1 results on the model trained on MOOD-IXTT1, only 0.38\% better). This not only shows that having more data is better but also the limits of the model on BraTS T1.

\begin{figure}
        \centering
        \includegraphics [width=0.49\textwidth] {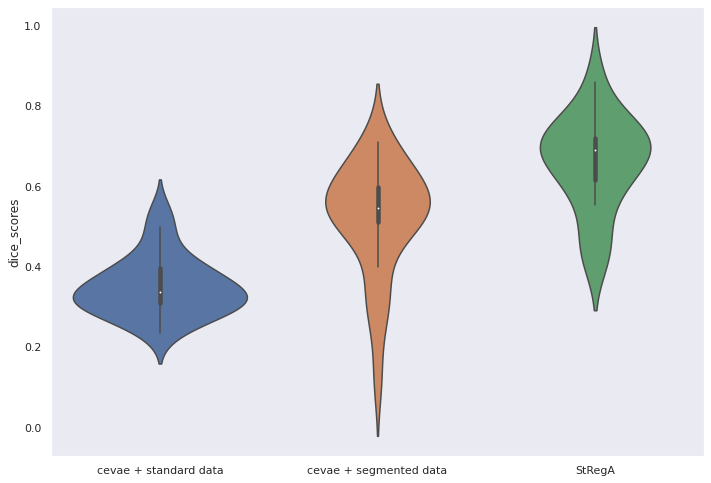}
        \caption{Violin plot comparing predictions on BraTS T2 data} 
        \label{fig:violin_compare_res}
\end{figure}

\begin{figure*}
        \centering
        \includegraphics [width=\textwidth] {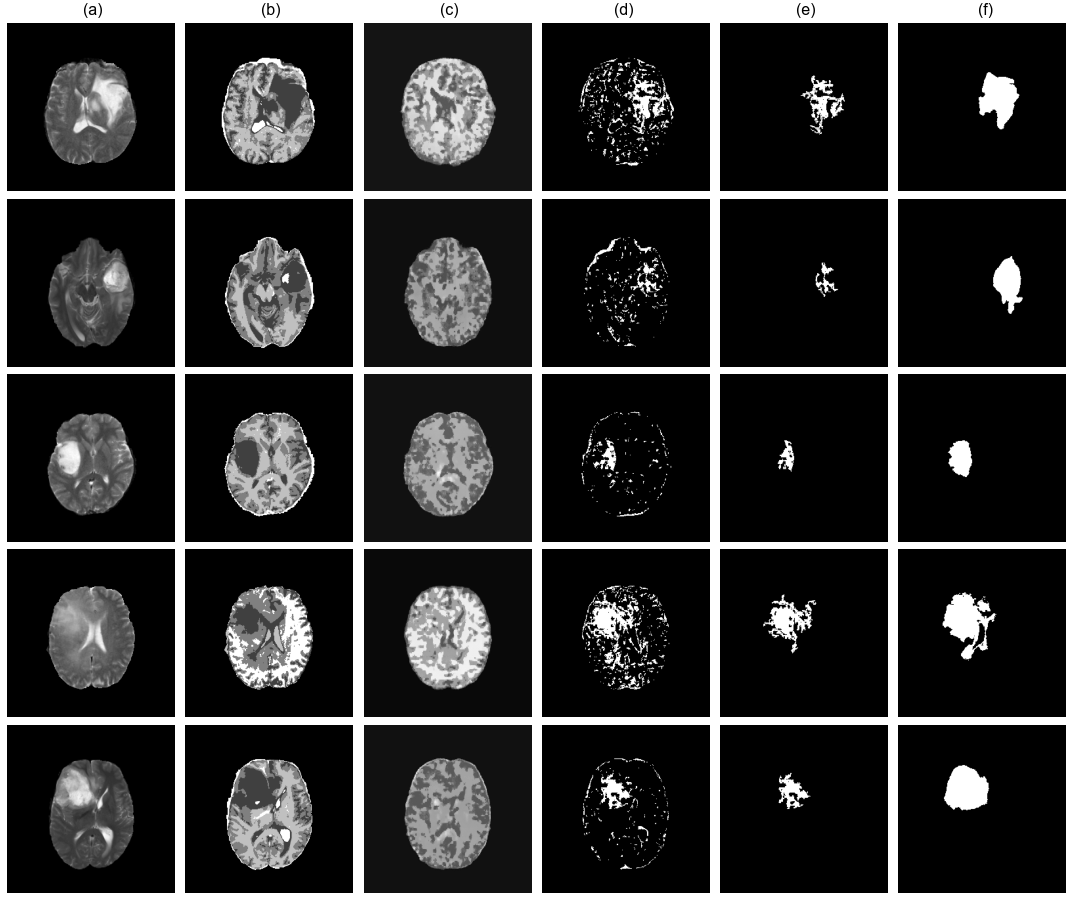}
        \caption{StRegA results on BraTS dataset (T2) images (a) Input Image (b) Segmented Image (c) Reconstruction (d) Manual and Otsu thresholding (e) Morphological Opening \textbf{Final prediction} (f) Ground Truth} 
        \label{fig:t2-final}
\end{figure*}

It can be observed that all the models perform better in the case of T2 than on T1ce. This can also be either because the tumours are easier to detect in T2 images than on T1ce images and could also be because of the segmentation performance of FSL in case of anomalous data, as FSL is typically used for anomaly-free images. Moreover, Fig.~\ref{fig:violin_compare_res} compares the range of resulting Dice scores for the task of anomaly detection in BraTS T2 images for ceVAE, ceVAE with StRegA preprocessing, and StRegA, with the help of violin plots. The white dot inside each of the violins represents the median Dice score, the thick black bar at the centre exhibits the interquartile range, and the thin dark line depicts the rest of the distribution excluding outliers. It can be seen that the segmentation assists the ceVAE model in achieving better scores, while the complete StRegA pipeline manages to improve the scores even further. StRegA results in a higher overall median Dice score than both the other methods, while its worst outlier is still better than the worst outliers of the other methods. Finally, Figures~\ref{fig:t1-final}~and~\ref{fig:t2-final} show the output of the different stages of the pipeline for BraTS T1 and T2, respectively. Similarly, Fig.~\ref{fig:syn-final} shows the stages for synthetic anomalies. 

\subsubsection{StRegA as an Assistive Technology}
\begin{figure}
    \centering
    \includegraphics[width=0.49\textwidth] {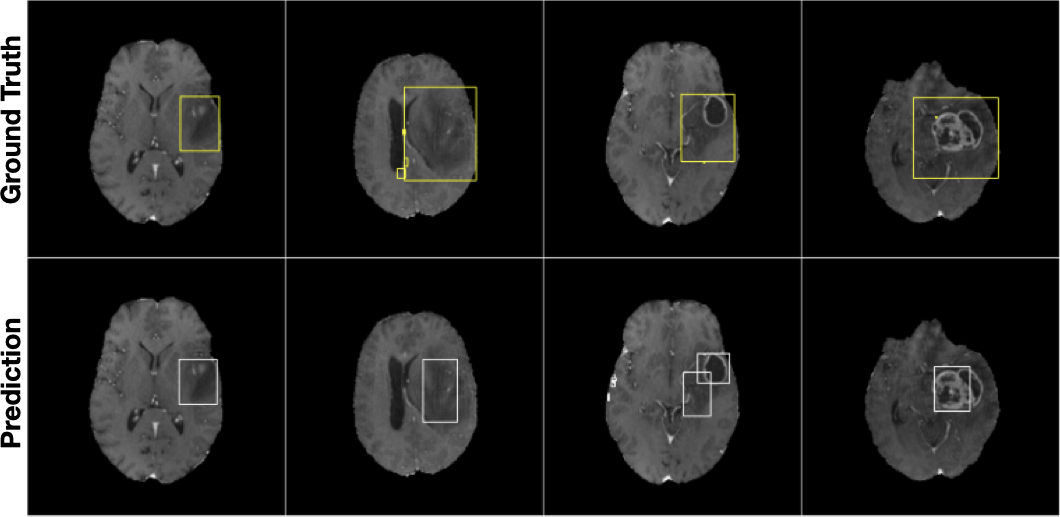}
    \caption{Anomaly localisation : BraTS T1ce}
    \label{fig:anomaly_locallise_T1}
\end{figure}

\begin{figure}
    \centering
    \includegraphics[width=0.49\textwidth] {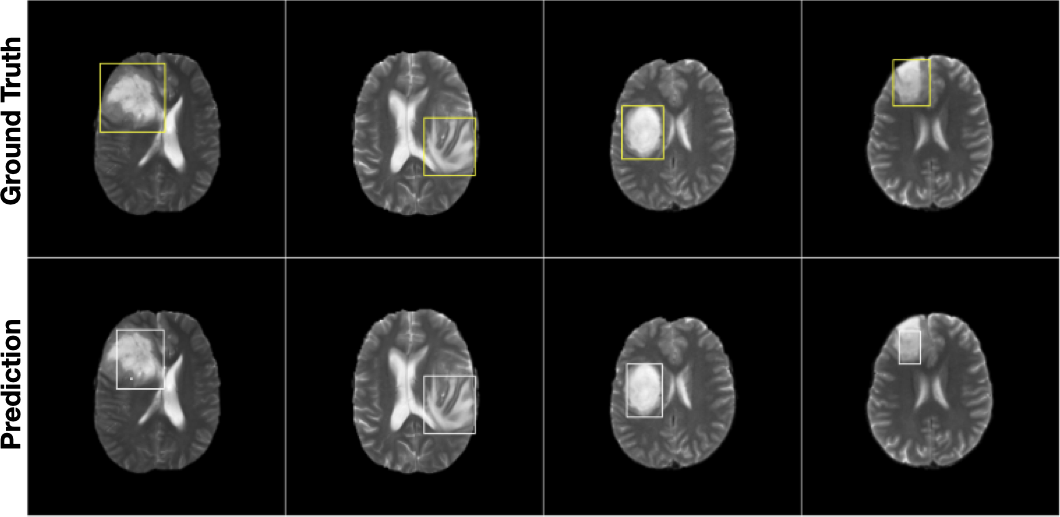}
    \caption{Anomaly localisation : BraTS T2}
    \label{fig:anomaly_locallise_T2}
\end{figure}

The earlier comparisons have demonstrated the proposed pipeline's superiority over the baseline models. However, several cases of under-segmentation were observed. Detecting if there is any anomaly or providing the location of the anomaly might also be sufficient to improve the clinical workflow by assisting the radiologists in filtering anomalous cases and getting an approximate location of the same. Figures.~\ref{fig:anomaly_locallise_T1}~and~\ref{fig:anomaly_locallise_T2} show the potential of using StRegA as an assistive technology to localise the anomalies in the MRIs for BraTS T1ce and T2, respectively. This was achieved by creating bounding boxes around the prediction of StRegA, and they were compared against the bounding boxes generated from the ground-truth labels. It can be observed that the proposed method was able to properly locate the anomalies. As seen earlier, the T2 results were better than the T1ce ones. For T1ce, under-estimations were observed most of the time. However, for T2, the results are better, and most of the time, StRegA managed to mark them appropriately. Despite the under-estimations, the results show the potential of using this method as assistive technology - as a decision support system.

\section{Discussion}
\label{sec:discussion}
The results demonstrate that the proposed StRegA pipeline achieved significantly better scores than the baseline models in detecting brain tumours and artificial anomalies while being trained on anomaly-free brain MRIs. The proposed pipeline aims at simplifying the input to the model by segmenting it. This same step also helps the model to generalise better in terms of the changes in contrast - which can be seen from the results of the models on the BraTS dataset. MOOD toy and the synthetic anomalous datasets had the same image contrast and resolution as the training dataset - on which the baseline models achieved comparable results, while BraTS had considerably different contrast - on which the baseline models struggled the most. Due to the fact that the input is simpler, a less-complex model has been seen to be sufficient (i.e. cceVAE instead of ceVAE). This helps in reducing the memory requirement - making it possible to work with more commonly-available hardware, and also reduces the chances of overfitting as the model has less number of trainable parameters.

Even though StRegA has performed significantly better than the baseline models, it is to be noted that the predictions are not perfect - which is commonly seen with UAD techniques, and there is a strong dependency on the preprocessing steps, as can be seen from the results of ceVAE and ceVAE with StrRegA preprocessing. It is also worth mentioning as FSL segmentation is typically used on anomaly-free images, using them on anomalous data might have induced some bias in the training-testing of the models which is not possible to quantify. Moreover, it was observed that the compact version of ceVAE proposed here performed better than the original ceVAE model, which might be attributed to the fact that the preprocessed images require a much simpler representation than images without preprocessing - consequently requiring less number of trainable parameters to learn the distribution of healthy-brain. Another interesting observation can that can be made from the results is that the models performed 21\% better on T2w images than T1ce from the BraTS dataset, even while being trained to focus on T1w images (T1ce was tested on models trained on purely T1w images, T2s were tested on models trained on a combination of T1 and T2). This can be because the tumours are easier to distinguish in T2 than in T1ce due to the significant intensity difference between tumour and non-tumour tissues. It can also be due to some bias induced by the FSL's segmentation. 

\begin{figure*}
        \centering
        \includegraphics [width=0.8\textwidth] {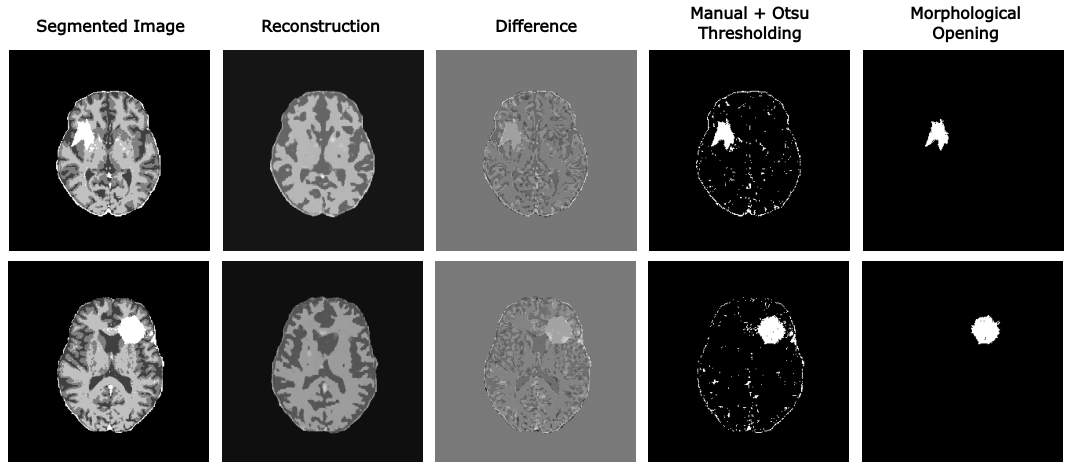}
        \caption{StRegA results on synthetically generated data} 
        \label{fig:syn-final}
\end{figure*}

\begin{figure*}
        \centering
        \includegraphics [width=\textwidth] {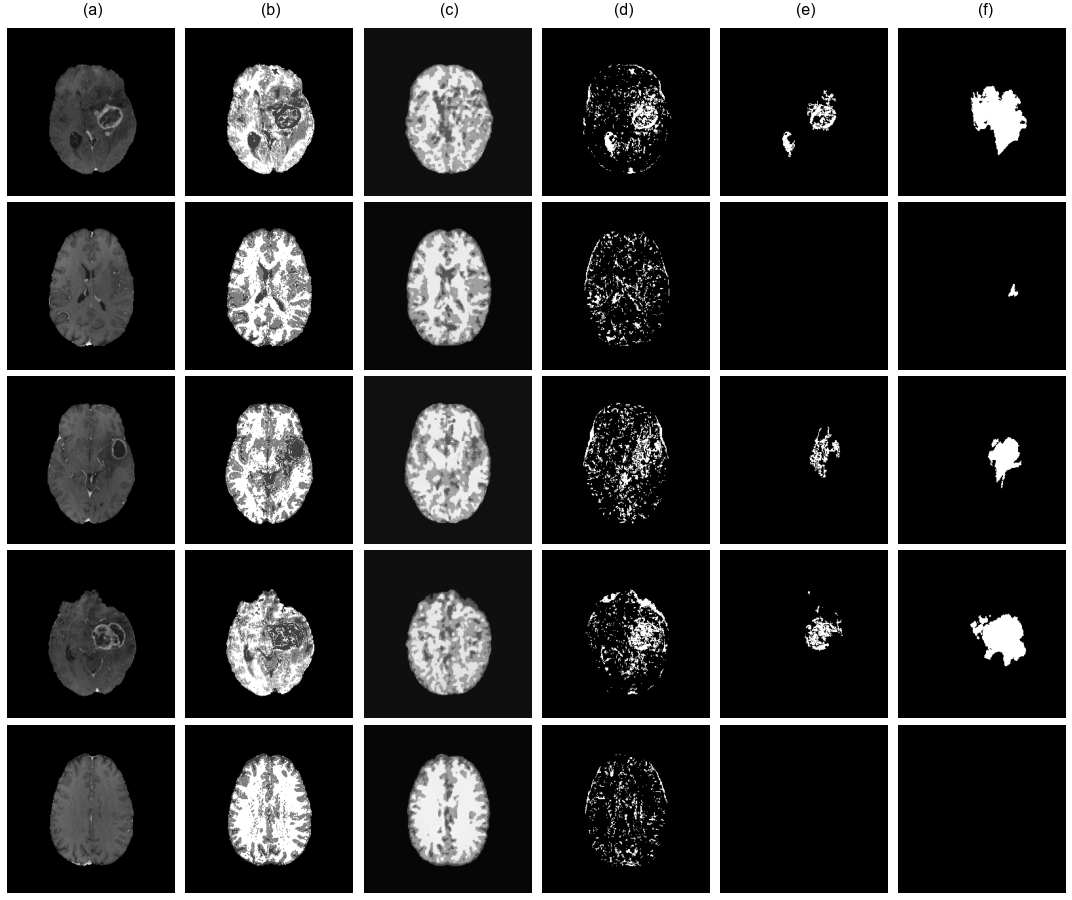}
        \caption{StRegA results on BraTS dataset (T1) images (a) Input Image (b) Segmented Image (c) Reconstruction (d) Manual and Otsu thresholding (e) Morphological Opening -> \textbf{Final prediction} (f) Ground Truth} 
        \label{fig:t1-final}
\end{figure*}

Undersegmentations can be observed in all the examples. This also resulted in a complete disappearance of a small anomaly as can be seen in Fig.~\ref{fig:t1-final}~(row 2). Hence, one immediate point of concern is the size of the anomaly in the input image. As morphological opening would eliminate the prediction if it is too small, the StRegA postprocessing steps might not be ideal for detecting small anomalies, e.g. multiple sclerosis. One way to deal with the small anomalies would be to reduce the area threshold. For synthetically generated anomalies, the pipeline performed almost perfectly, as can be seen in Fig.~\ref{fig:syn-final}. Nevertheless, these models were trained in an unsupervised fashion – without any labelled training data. Hence, the requirement of large manually annotated training data can be mitigated. The anomalies detected by such a method can also be further used as weak labels to train weakly-supervised or semi-supervised models - which might improve the final segmentation quality without requiring much manual intervention.   

StRegA, similar to the baselines experimented here, works in 2D and relay upon the model's reconstruction error. This way of detecting anomalies is suitable for local structural anomalies, such as tumours, but might not be suitable for detecting changes in brain atrophy, such as the change in brain atrophy due to the Alzheimer's disease. Going for a 3D method and extending the way of detecting the anomalies might help the method detect such anomalies. It is worth mentioning that the preprocessing techniques like brain extraction and segmentation used in this approach aiming at simplifying the input provided to the model are domain-specific for brain MRIs, while augmentations like random bias field artefacts might only be realistic for MRIs. However, the principal idea proposed here - simplifying the input before supplying to a VAE-based model for anomaly detection, as well as the cceVAE model - is applicable to other domains and applications, such as liver tumour detection in CT or MRI~\cite{gul2022deep}, lesion detection in chest x-rays~\citep{qi2022directional}, even for non-medical domains like fault-detection in additive manufacturing~\citep{iuso2022segment}. For the same, different domain-specific preprocessing steps (e.g. segmenting abdominal MRI input into different organ classes - similar to the FSL segmentation performed here) to simplify the input have to be ascertained and evaluated.

As a final note, StRegA as an assistive technology might help radiologists notice anomalies that might otherwise be overlooked. Diagnostics by radiologists remain undiscovered in up to 5-10 \% of cases \citep{bruno2015understanding}. The cause of these errors is due to unintentional bias, where radiologists focus more on the centre part of the MRI while neglecting peripheral findings. The ever-increasing number of radiological images for assessment increases the workload in radiology which in turn can result in more missed brain pathologies. The presented method might assist in localising anomalies (as seen in Fig.~\ref{fig:anomaly_locallise_T2}~and~\ref{fig:anomaly_locallise_T2}) and improve the overall diagnostic accuracy. 

\section{Conclusions and Future Work}
\label{sec:conclusions}
This paper address the challenge of anomaly detection in an unsupervised manner by presenting an unsupervised anomaly detection system named StRegA: \textbf{S}egmen\textbf{t}ation \textbf{Reg}ularised \textbf{A}nomaly, which combines a modified and compact version of the ceVAE model (cceVAE) with pre- and postprocessing steps. The proposed method was trained using anomaly-free brain MRIs from two benchmark datasets and was then evaluated for the task of detection of brain tumours. The proposed method was compared against four baselines and outperformed all of them with statistical significance. StRegA achieved Dice scores of 0.531±0.112 and 0.642±0.101 while segmenting tumours, as well as 0.723±0.134 and 0.859±0.112 while segmenting artificial anomalies, from T1w and T2w MRIs, respectively - resulting in 49\%, 83\%, 5\%, and 21\% improvements, respectively in the four scenarios, over the best performing baseline model ceVAE, while achieving 2\%, 23\%, 2\%, and 10\% improvements over ceVAE with StRegA preprocessing. Undersegmentations were observed in most cases, however, the method could localise the anomalies properly. This work has shown the potential of using this method as part of a decision support system. 

In the future, this method will be evaluated for the tasks of detecting other types of anomalies than brain tumours - to evaluate its generalisability to other pathologies. 
The current method faces difficulties while detecting small anomalies - this might be addressed by modifying the postprocessing steps. Furthermore, techniques other than segmentation to simplify the input to the model will also be evaluated in the near future.

\section*{Acknowledgement}
This work was in part conducted within the context of the International Graduate School MEMoRIAL at Otto von Guericke
University (OVGU) Magdeburg, Germany, kindly supported by the European Structural and Investment Funds (ESF) under the
programme "Sachsen-Anhalt WISSENSCHAFT Internationalisierung" (project no. ZS/2016/08/80646) and was supported by the federal state of Saxony-Anhalt (“I 88”). 

\begin{table*}
  \begin{center}
    \caption{Comparative results of StRegA with other models on T1 data.}
    \label{tab:t1comparenumbers}
    \begin{tabular}{ |c|c|c|c|c|}
    \hline
    \multirow{2}{*}{\bfseries Model} & 
    \multicolumn{3}{|c|}{\bfseries Dice Score ($\mu\pm\sigma$)} \\
    \cline{2-4} 
     & MOOD Toy Data & Synthetic Anomalous Data &  BraTS T1w \\
    \hline
    SkipAE & 0.443 $\pm$ 0.136 & 0.498 $\pm$ 0.164 & 0.209 $\pm$ 0.113\\
    \hline
    GMVAE & 0.657 $\pm$ 0.121 & 0.711 $\pm$ 0.105 & 0.337 $\pm$ 0.124\\
    \hline
     ceVAE & 0.633 $\pm$ 0.145 & 0.690 $\pm$ 0.164 & 0.356 $\pm$ 0.132\\
    \hline
     ceVAE (with StRegA preprocessing) & 0.724 $\pm$ 0.132 & 0.710 $\pm$ 0.111 & 0.520 $\pm$ 0.108\\
    \hline
    \textbf{StRegA}  & 0.737 $\pm$ 0.098 & 0.723 $\pm$ 0.134 & 0.531 $\pm$ 0.112\\
    \hline
    \end{tabular}
  \end{center}
\end{table*}
\begin{table*}
  \begin{center}
    \caption{Comparative results of StRegA with other models on T2w data.}
    \label{tab:t2comparenumbers}
    \begin{tabular}{ |c|c|c|c|c|}
    \hline
    \multirow{2}{*}{\bfseries Model} & 
    \multicolumn{3}{|c|}{\bfseries Dice Score ($\mu\pm\sigma$)} \\
    \cline{2-4} 
     & MOOD Toy Data & Synthetic Anomalous Data &  BraTS T2 \\
    \hline
    SkipAE & 0.517 $\pm$ 0.145 & 0.498 $\pm$ 0.173 & 0.221 $\pm$ 0.101\\
    \hline
    GMVAE & 0.699 $\pm$ 0.121 & 0.701 $\pm$ 0.121 & 0.340 $\pm$ 0.167\\
    \hline
     ceVAE & 0.724 $\pm$ 0.145 & 0.712 $\pm$ 0.132 & 0.350 $\pm$ 0.071\\
    \hline
     ceVAE (with StRegA preprocessing) & 0.797 $\pm$ 0.099 & 0.783 $\pm$ 0.111 & 0.522 $\pm$ 0.135\\
    \hline
    \textbf{StRegA}  & 0.856 $\pm$ 0.098 & 0.859 $\pm$ 0.112 & 0.642 $\pm$ 0.101\\
    \hline
    \end{tabular}
  \end{center}
\end{table*}


\bibliography{mybibfile}

\end{document}